\documentclass[prb,twocolumn]{revtex4}
\usepackage{graphicx}

\begin{document}
\title{Upper limit on spontaneous supercurrents in Sr$_2$RuO$_4$}
\author{
J.R. Kirtley,$^{1,2,3}$
C. Kallin,$^4$
C.W. Hicks,$^1$
E.-A. Kim,$^{5,6}$
Y. Liu,$^{7}$
K.A. Moler,$^{1,5}$
Y. Maeno,$^8$
and
K.D. Nelson,$^7$
}

\affiliation{$^1$ Department of Applied Physics, Stanford University, Palo Alto, CA, USA}
\affiliation{$^{2}$ IBM Watson Research Center, Yorktown Heights, NY, USA}
\affiliation{$^{3}$ Faculty of Science and Technology
and MESA$^+$ Institute for Nanotechnology, University of Twente, Enschede, The Netherlands}
\affiliation{$^4$ Department of Physics and Astronomy, McMaster University,
Hamilton, ON L8S 4M1, Canada}
\affiliation{$^5$ Department of Physics, Stanford University, Palo Alto, CA, USA}
\affiliation{$^6$ Stanford Institute of Theoretical Physics, Stanford University,
Palo Alto, CA, USA}
\affiliation{$^7$ Department of Physics, Pennsylvania State University, University Park,
PA, 16802, USA}
\affiliation{$^8$ Department of Physics, Kyoto University, Kyoto 606-8502, Japan}

\date{\today}

\begin{abstract}
It is widely believed that the perovskite Sr$_2$RuO$_4$ is an unconventional
superconductor with broken time reversal symmetry. It has been predicted that superconductors
with broken time reversal symmetry should have spontaneously generated supercurrents at
edges and domain walls. We have done careful imaging
of the magnetic fields above Sr$_2$RuO$_4$ single crystals using scanning Hall bar and SQUID microscopies,
and see no evidence for such spontaneously generated supercurrents. We use the results from our
magnetic imaging to place upper limits on the spontaneously generated supercurrents at edges and domain walls
as a function of domain size.  For a single domain, this
upper limit is below the predicted signal by two orders of magnitude.
We speculate on the causes and implications of the lack of large spontaneous
supercurrents in this very interesting superconducting system.
\end{abstract}

\maketitle



\section{Introduction}
The perovskite superconductor Sr$_2$RuO$_4$\cite{maeno1994} (T$_c$=1.5K)
is believed for a number of reasons
to have unconventional pairing symmetry.\cite{mackenzie2003}
Muon spin resonance experiments are consistent with the generation of large but sparse internal
magnetic fields when
Sr$_2$RuO$_4$ becomes superconducting, indicating a superconducting state with broken time reversal
symmetry.\cite{luke1998} This finding is supported by the observation of the onset at the superconducting
transition temperature of a Kerr effect rotation of light polarization upon reflection,\cite{xia2006}
consistent with large ($\sim$50-100$\mu$m) domains with broken chiral symmetry. The first phase sensitive
Josephson tunneling measurements suggested a static $\pi$-phase shift between opposite faces
of a Sr$_2$RuO$_4$ single crystal,\cite{nelson2004} a result obtained only when
the sample was prepared by controlled slow cooling, implying that the sample
could be made to possess a small even number of domain walls separating the opposite faces.
On the other hand, more recent Josephsen tunneling measurements point
toward small ($\sim$1$\mu$m)
dynamic order parameter
domains.\cite{kidwingara2006}
Magnetic imaging of the $ab$ face of a single crystal of Sr$_2$RuO$_4$ using a micron sized
SQUID shows vortex coalescence on a scale of $\sim$10 $\mu$m that may be related to
a domain structure in the superconducting order parameter.\cite{dolocan2005}
These observations, as well as others, have been interpreted
in terms of a superconducting
order parameter with spin triplet,\cite{nmr} chiral $p_x \pm ip_y$ Cooper
pairing symmetry.  However, Hall bar microscopy measurements\cite{bjornsson2005} did not
observe the magnetic fields expected at the surface or edges of a
superconductor with broken chiral symmetry.  Here, we report on new
scanning SQUID microscope measurements, and we also further analyze some of
the Hall bar
measurements, showing that, if they exist, the local fields at the surfaces and edges of Sr$_2$RuO$_4$
single crystals are much smaller than those expected at the surfaces and edges of a
chiral p-wave superconductor.

Although
$p_x \pm i p_y$ pairing symmetry is fully gapped, specific heat, \cite{nishizaki2000}
nuclear relaxation rate, \cite{ishida2000}
and thermal conductivity measurements \cite{tanatar2001a,tanatar2001b,izawa2001}
all show a power law temperature dependence, suggesting
the presence of line nodes. Among other suggestions, one possibility is that the $\gamma$ band, the
band with the primary contribution
to superconductivity in Sr$_2$RuO$_4$,  has nodeless $p_x \pm i p_y$ pairing symmetry, but induces
superconductivity with a line of nodes in the other ($\alpha$ and $\beta$) bands. \cite{zhitomirsky2001}
Anisotropy in the gap function \cite{miyake1999} has been supported experimentally
by specific heat \cite{deguchi2004a,deguchi2004b} and ultrasound attenuation \cite{lupien2001}
measurements.

The issue of the broken time reversal symmetry in the superconducting state
of Sr$_2$RuO$_4$, aside from intrinsic interest, has taken
on new urgency with several proposals for error tolerant quantum logic elements taking
advantage of this property. \cite{zagoskin2003a,zagoskin2003b,dassarma2006}

There are useful analogies between a chiral $p_x \pm ip_y$ superconductor and
a ferromagnet.\cite{kidwingara2006,rice2006}
A single domain ferromagnet has a uniform magnetization which is equivalent
to the field produced by a current sheet circulating around the surface,
in the appropriate geometry, while a single domain
$p_x + ip_y$ superconductor carries
an intrinsic angular momentum of $\hbar$ per Cooper pair,\cite{kita1997} which one
would expect to lead to
an actual surface current sheet, confined within a healing length proportional to the
coherence length of the surface.\cite{stone2004}
However, the field generated by this current
must be screened inside the superconductor by a diamagnetic shielding
current flowing
within the penetration depth of the surface, so that $B=0$ inside the
superconductor.  The net result is a spontaneous
magnetization within the healing length plus the penetration depth of the sample edges,
which is greatly reduced from that expected from the simple
ferromagnetic analogy, but which is still substantial and, using
parameters appropriate to Sr$_2$RuO$_4$, predicted to give
rise to local fields as large as 1mT under certain
assumptions.\cite{matsumoto1999,kwon2001}  The superconductor can also
support domains in which regions of $p_x + ip_y$ coexist with regions
of $p_x - ip_y$ order.  Although the net magnetization vanishes at the
boundaries between such domains, the local fields, which extend over
the penetration depth on either side of the wall, can be as large as
2mT.\cite{matsumoto1999,kwon2001} Other than direct phase sensitive
measurements, the detection of such fields would
be one of most direct confirmations of a superconducting order
parameter with
time reversal symmetry breaking since the spontaneous boundary and domain wall
supercurrents are expected by symmetry\cite{sigrist1991} and would
have no other obvious
explanation.  It is therefore appropriate to attempt to image the magnetic fields arising from these
spontaneous supercurrents using scanning magnetic microscopy.

\section{Magnetic imaging}

We have performed scanning magnetic imaging of the $ab$ and $ac$ faces of single crystals of
Sr$_2$RuO$_4$. The magnetic
images reported here were made at Stanford
with a dilution refrigerator based Hall bar/SQUID
microscope \cite{bjornsson2001} with a base temperature below 100mK, and at IBM with
a $^3$He based
scanning SQUID microscope with a base temperature below 300mK. Our SQUID sensors had square
pickup loops 8 $\mu$m on a side; the Hall bars had roughly square
effective areas 0.5 $\mu$m on a side. The Hall bar measurements were made in a residual field
of about 2.5$\mu$T; the SQUID measurements were made in a residual field of 75nT,
compensated for fields perpendicular to the scanning direction to less than 10nT using a
small Helmholtz coil. The SQUID measurements were made after cooling the samples through
the superconducting transition temperature at a rate of about 1mK/sec.
Some of the Hall bar data discussed in this paper
has been reported previously. \cite{bjornsson2005} However, here we
make a more quantitative comparison of this data with theory.

The Sr$_2$RuO$_4$ single crystals used in our
experiments were grown using a floating zone method.\cite{mao2000}
The samples used for the IBM SQUID measurements were mounted in epoxy
and polished so that either the $ab$ or $ac$ face was part of a smooth plane, allowing
scanning across the edges of the crystal.\cite{nelson2004} Some of the samples used for SQUID
microscopy were the
same as for phase sensitive experiments on the pairing symmetry of Sr$_2$RuO$_4$,
\cite{nelson2004} and had layers of SiO and Au$_{0.5}$In$_{0.5}$ (T$_c$= 0.4-0.5K) deposited on some of the crystal
faces perpendicular to the scanned face. These additional layers should have had no effect on
the magnetic imaging experiments reported here. The critical temperature of the crystals were
measured to be $>$ 1.4K using scanning and bulk susceptometry measurements.

\begin{figure}
\includegraphics[width=3.5in]{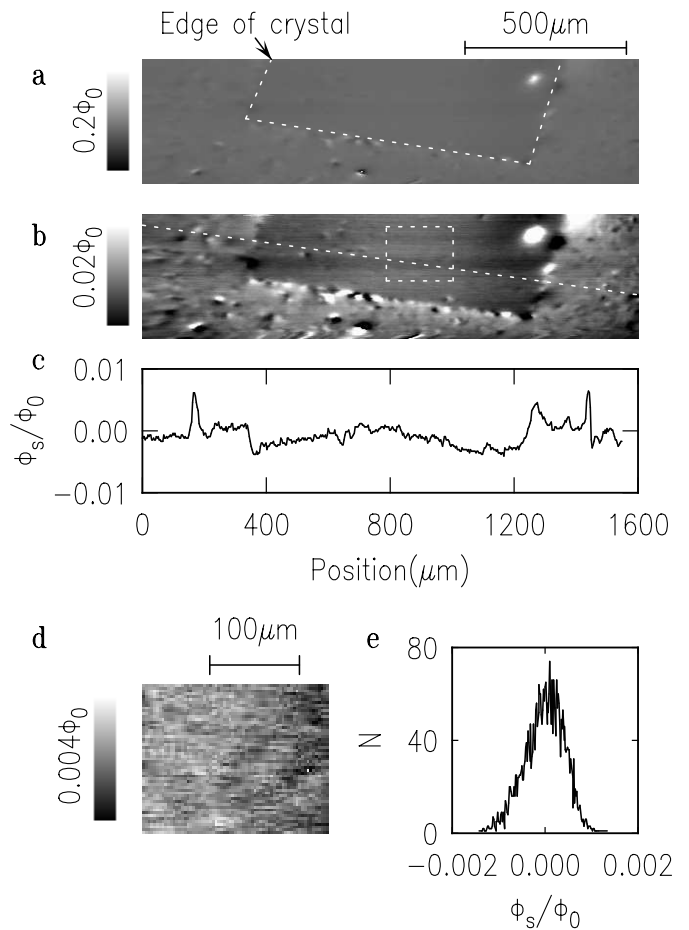}
\vspace{0.1in}
\caption{SQUID microscope image of the $ab$ face of a Sr$_2$RuO$_4$ single crystal,
cooled in a field $B_z<$10 nT and imaged at T=0.27K with
an 8 $\mu$m square pickup loop. {\bf a} Pseudocolor image with full-scale variation of 0.2 $\Phi_0$
($\Phi_0 = h/2e$) in magnetic flux through the SQUID pickup loop. The dashed line in {\bf a} shows the
outlines of the crystal.
{\bf b} Same image as {\bf a},
but with the pseudocolor scale expanded to 0.02 $\Phi_0$.
The dashed line in {\bf b} shows the line traced by the cross-section
in {\bf c}. The dashed rectangle in {\bf b} shows the area of the image expanded in {\bf d}. {\bf e}
is a histogram of pixel values for the data displayed in {\bf d}.
}
\label{fig:abimage}
\end{figure}

\begin{figure}
\includegraphics[width=3.5in]{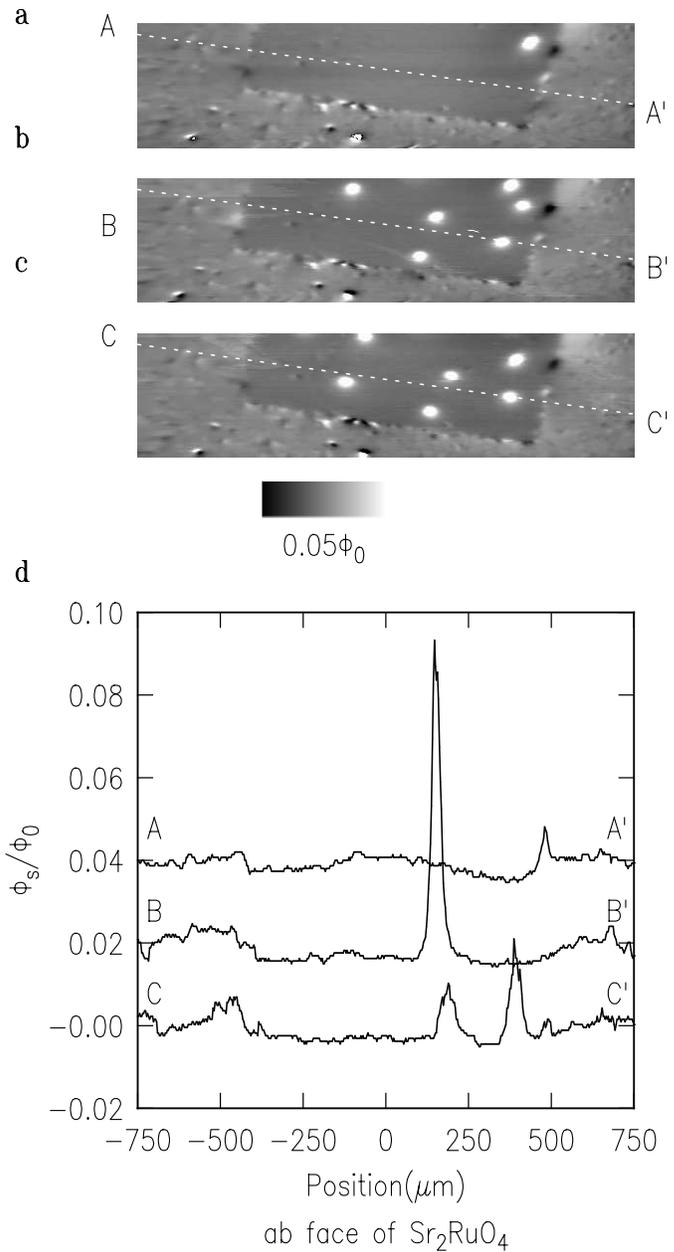}
\vspace{0.1in}
\caption{Comparison of SQUID microscope images of an $ab$ face of a Sr$_2$RuO$_4$ crystal
after three different cooldowns in slightly different magnetic fields.}
\label{fig:multcrs}
\end{figure}

Figure \ref{fig:abimage} shows a SQUID microscope image of the $ab$ face of a Sr$_2$RuO$_4$ single
crystal. The largest feature evident in this image
(Fig. \ref{fig:abimage}a) is an isolated Abrikosov vortex. When the pseudocolor scale
is expanded to $\Delta \Phi_s = 0.02 \Phi_0$ (Fig. \ref{fig:abimage}b)
magnetic features become apparent in the epoxy and along
the edges of the crystal. We believe these features are not due to
the superconductivity of the Sr$_2$RuO$_4$ because
they are unchanged from cooldown to cooldown in different fields.
Figure \ref{fig:multcrs} compares images from 3 different cooldowns of the same crystal, in
nominal ambient plus compensating fields perpendicular to the scanning plane of zero (Fig. \ref{fig:multcrs}{\bf a}),
10nT (Fig. \ref{fig:multcrs}{\bf b}), and 15nT (Fig. \ref{fig:multcrs}{\bf c}).
The number and positions of the Abrikosov vortices in the top (ab) face of the
crystal, and an interlayer vortex emerging from the left (ac face) edge
(Fig. \ref{fig:multcrs}{\bf b}) of the crystal
change from cooldown to
cooldown, but the sharp features at the edge of the sample are remarkably reproducible.
These edge features may be the result of the polishing process, such as topographical
or magnetic features from particles trapped in the epoxy. Note that features very similar to the edge
features are apparent in the epoxy far from the sample edge. Above the sample itself the
flux image is relatively smooth, with a broad background (Fig. \ref{fig:abimage}c).
We believe that this broad background is the
result of magnetic flux coupled into the SQUID through sections outside of the pickup loop.
A clear demonstration of this effect appears in Ref. \onlinecite{kirtley1998}. On top of the broad background, two steps in the cross-section
(Fig. \ref{fig:abimage}c) correspond to the edges of the crystal. We believe that these steps are due
to small supercurrents circulating around the entire sample due to uncompensated residual fields (see Figure \ref{fig:srocrs}).
Figure \ref{fig:abimage}d shows a magnified image of a section of the crystal (indicated by the
box in Fig. \ref{fig:abimage}b), with no magnetic features larger than a few m$\Phi_0$ over an
area of several hundred
microns on a side.

\begin{figure}
\includegraphics[width=3.5in]{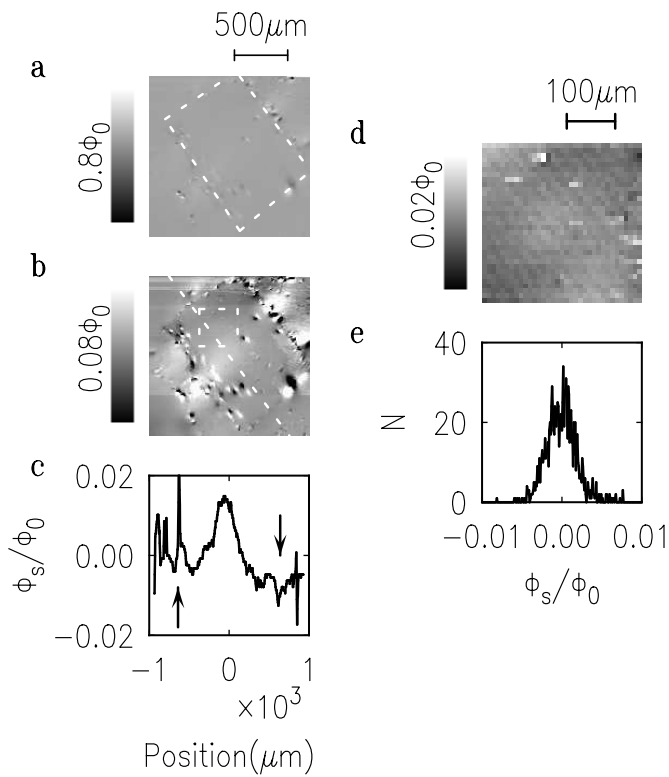}
\vspace{0.1in}
\caption{SQUID microscope image of the $ac$ face of a Sr$_2$RuO$_4$ single crystal
cooled in nominally zero field and imaged at T=0.27K with
an 8 $\mu$m square pickup loop. {\bf a} Pseudocolor image
$\Delta \Phi_s = 0.8 \Phi_0$.
The dashed line in {\bf a} shows the positions of the outer edges of the crystal.
{\bf b} Same image as {\bf a},
but with $\Delta \Phi_s = 0.08 \Phi_0$
A few interlayer vortices with both positive and negative signs
are visible near the lower-left edge of the crystal.
The dashed line in {\bf b} is along the $a$-axis and shows the data traced by the cross-section
in {\bf c}. The arrows in {\bf c} indicate the edges of the crystal.
The dashed square in {\bf b} shows the area of the image expanded in {\bf d}.
The diagonal stripes visible in {\bf d} are due to 60 Hz noise.
{\bf e}
is a histogram of pixel values for the data displayed in {\bf d}. }
\label{fig:acimage}
\end{figure}

Similar results were obtained when SQUID microscope images were taken of the $ac$ face of a
Sr$_2$RuO$_4$ single crystal (Figure \ref{fig:acimage}). In this case there were a number of
interlayer vortices with flux both emerging from and entering into the crystal surface near
the left edge of the crystal (Fig. \ref{fig:acimage}b). Just as for the $ab$ face,
there were sharp magnetic features along the edges of the crystal and in the epoxy which did not
appear to be correlated with the superconductivity of the Sr$_2$RuO$_4$, as well as broad magnetic
backgrounds, but sharp magnetic features were absent from large areas of the crystal face.

\begin{figure}
\includegraphics[width=3.5in]{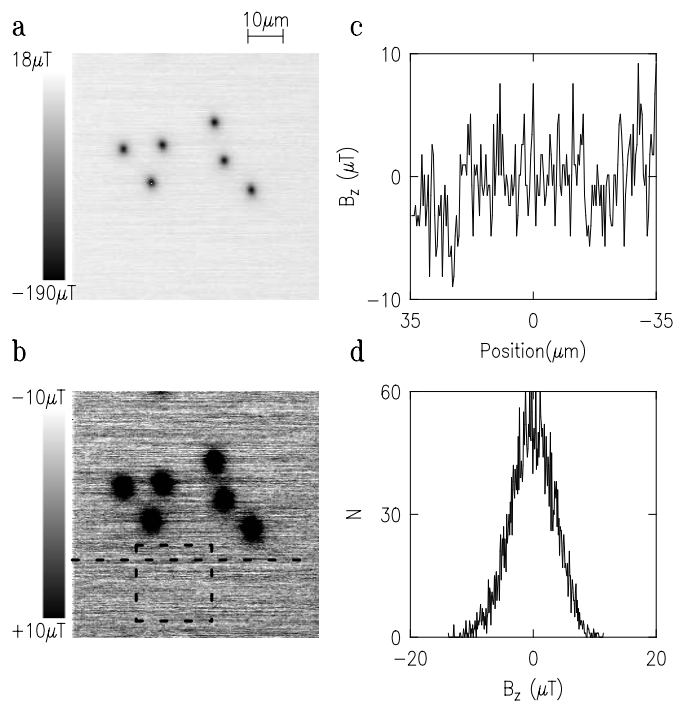}
\vspace{0.1in}
\caption{{\bf a} Scanning Hall bar image of $ab$ face of Sr$_2$RuO$_4$ single crystal, cooled in
$\sim 2.5 \mu$T and imaged at a temperature below 100mK using a Hall bar with a sensor area 0.5$\mu$m on a side.
In this image the mean of each scan line was subtracted from the raw data
to remove slow drift in the sensor Hall voltage.
{\bf b} Same area as {\bf a} but with an expanded pseudocolor scale.
The dashed line in {\bf b} shows the line traced by the data cross-section in {\bf c}.
The dashed square in {\bf b} shows the area for which a histogram of pixel inductance values are displayed in
{\bf d}.
}
\label{fig:hallimag}
\end{figure}

The samples used in the Hall bar measurements were cleaved. 1$\mu$m diameter,
$\sim$1 $\mu$m deep holes were milled on a 20 $\mu$m grid on the upper surface using
a focussed ion beam, to create artificial edges.\cite{bjornsson2005}
Figure \ref{fig:hallimag} shows a scanning Hall bar image of the $ab$ face, with a regular
array of 1 micron holes at a pitch of 20 microns, of a Sr$_2$RuO$_4$ single
crystal. There are a few Abrikosov vortices apparent in this image, but the area away from these vortices
is featureless.  In particular, no features were observed in connection with the edges or
interiors of the 1 micron holes. Since the 1 $\mu$m deep holes did not even act as effective
pinning centers for the vortices, they may not have served as significant singularities
to create edge currents.
The outer edges of the crystal were not scanned in the Hall bar
measurements.

\section{Modeling}

Matsumoto and Sigrist \cite{matsumoto1999} (MS) have solved the Bogoliubov-De Gennes
equations using
a quasi-classical approximation for the cases of an edge between a semi-infinite,
ideal $p_x +i p_y$ superconductor and vacuum, and a domain boundary between a $p_x + i p_y$
superconductor and a $p_x - i p_y$ superconductor.  Their solutions are fully self-consistent
so that they include the effect of screening currents.  They predict substantial supercurrents and
consequent magnetic fields spontaneously generated at edges and domain boundaries.
For example, the peak magnetic fields in these calculations
correspond to 1 mT for edges, and 2 mT for domain walls
using values for the coherence length ($\xi_0$=66 nm) and penetration depth ($\lambda_L$= 190 nm)
suitable for Sr$_2$RuO$_4$. However, some modeling is required to compare our experimental
results with the MS predictions because we measure the magnetic fields above
the surface, rather than inside the sample.

The simplest approach to this problem is to assume that the magnetic fields at the surface
of the sample are the same as those in the bulk. This neglects
field spreading and any change in superconducting shielding due to the finite sample geometry.
However, in our case
the size of the magnetic sensor and its spacing from the sample
are large relative to the coherence length and penetration depth, so that the field
averaging from these effects are larger than
the additional
effects of field spreading and changes in superconducting shielding.
The field averaging effects from finite sensor size and height can be shown rigorously to
be larger than field spreading, for example,
in the similar problem of vortex fields spreading from the surface of
a superconductor and imaged with a SQUID microscope.\cite{chang1992,kirtley1999}
 In the remainder of this section, we will
neglect changes in the currents near the surface due to the finite sample geometry.
We will show below that the effect of finite sample geometry only leads to suppression
of the expected signal by 30\% compared to what is expected from the edge currents
of an infinite sample.
The finite sample geometry effects for edge currents
are expected to be similar to those for domain walls and are also discussed in the following section on surface
screening effects.

It is well known \cite{roth1988} that if the normal component of the magnetic field $B_z(x,y,z)$
is known at all points of a surface $z=0$ the magnetic field in free space
at a height $z$ above that surface is given by
\begin{equation}
\tilde{B}_z(k_x,k_y,z)=\tilde{B}_z(k_x,k_y,z=0)e^{-kz},
\label{eq:wikswo_trick}
\end{equation}
where $\tilde{B}_z(k_y,k_y,z)$ is the 2-dimensional Fourier
transform of $B_z(x,y,z)$ and $k=\sqrt{k_x^2+k_y^2}$.
To model the magnetic signals in our experimental SQUID and Hall
bar microscope geometries, we assume a particular domain structure with the magnetic fields, $B_z$,
at each edge and domain boundary, at the surface $z=0$, taken to be those predicted by
Matsumoto and Sigrist \cite{matsumoto1999} for an infinite sample. We then propagate the fields
to a height $z$ using
Eq. \ref{eq:wikswo_trick}, integrate over an area appropriate for the SQUID or Hall bar sensor
to obtain a magnetic flux, and  divide by the area of
the sensor for the case of the Hall bar to get an average magnetic field.
We will refer to
this model as the ``extended-Matsumoto-Sigrist'' model to distinquish it both from
the prediction made by Matsumoto-Sigrist for an ideal (infinite) geometry and from the
more accurate model which includes additional screening effects due to the finite
geometry, as discussed in the next section.  The original Matsumoto-Sigrist results
are scaled in field by $B_c=\Phi_0/2\sqrt{2}\pi\xi_0\lambda_L$, where $\Phi_0=h/2e$ is the superconducting
flux quantum, $\xi_0$ is the coherence length and $\lambda_L$ is the London
penetration depth. For the modeling presented
here we take $\xi_0 = 66 nm$ and $\lambda_L = 190 nm$.\cite{mackenzie2003}

\begin{figure}
\includegraphics[width=3.5in]{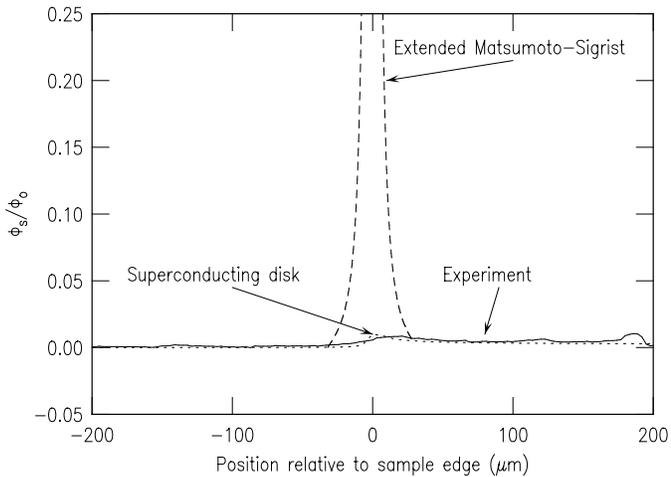}
\vspace{0.1in}
\caption{
Cross-section through the image of the $ab$ face of Sr$_2$CuO$_4$ displayed
in Figure \ref{fig:abimage} (solid line). The short-dashed line is the prediction for a superconducting disk
in a uniform residual field of 3 nT. The long-dashed line (with a peak at $\Phi_s/\Phi_0$ = 1.1)
is the prediction for a single domain
$p_x+ip_y$ superconductor of the extended-Matsumoto-Sigrist model as described in the text, assuming
a square pickup loop 8$\mu$m on a side, at a height of 3$\mu$m above the sample.
Here the superconductor is positioned to the left of 0$\mu$m, with epoxy to the right.
}
\label{fig:srocrs}
\end{figure}

Figure \ref{fig:srocrs} compares the results of this calculation (long-dashed line) with the
experimental cross-section
of the image shown in Figure \ref{fig:abimage} (solid line).
Also shown for comparision is the predicted
cross-section for an ideal superconducting disk in a uniform residual field\cite{jackson} of
3 nT.  The small steps in flux
at the edges of the crystals in Figures \ref{fig:abimage} and \ref{fig:acimage} can be attributed
to shielding of a very small residual background field. These steps are much smaller than the
peaks predicted by the extended-Matsumoto-Sigrist model for a single domain.

\begin{figure}
\includegraphics[width=3.5in]{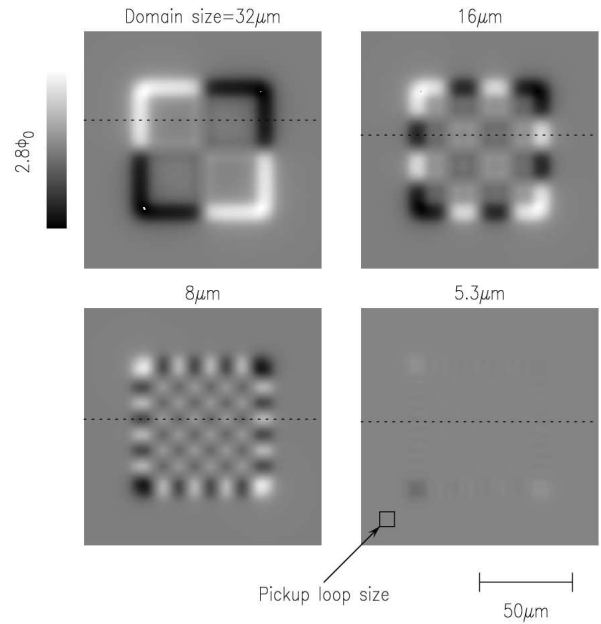}
\vspace{0.1in}
\caption{
Predicted magnetic fluxes through an 8 $\mu$m square pickup loop, 3 $\mu$m above the sample surface,
for a 64 $\mu$m square $p_x \pm p_y$ superconductor with various domain sizes, using the predictions
for the edge and domain wall currents of Matsumoto and Sigrist as
described in the text. The dashed lines
in the figure show the positions of the cross-sections displayed in Fig. \ref{fig:fluxncrs}.
}
\label{fig:fluxndom}
\end{figure}

\begin{figure}
\includegraphics[width=3.5in]{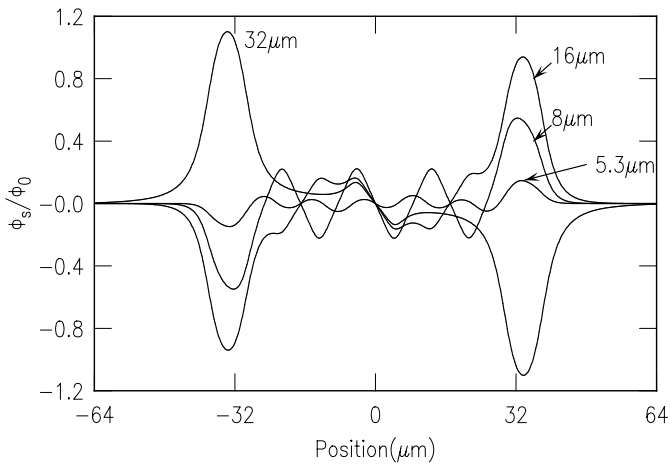}
\vspace{0.1in}
\caption{
Cross-sections through the modeling images of Fig. \ref{fig:fluxndom} for various domain sizes.
}
\label{fig:fluxncrs}
\end{figure}

\begin{figure}
\includegraphics[width=3.5in]{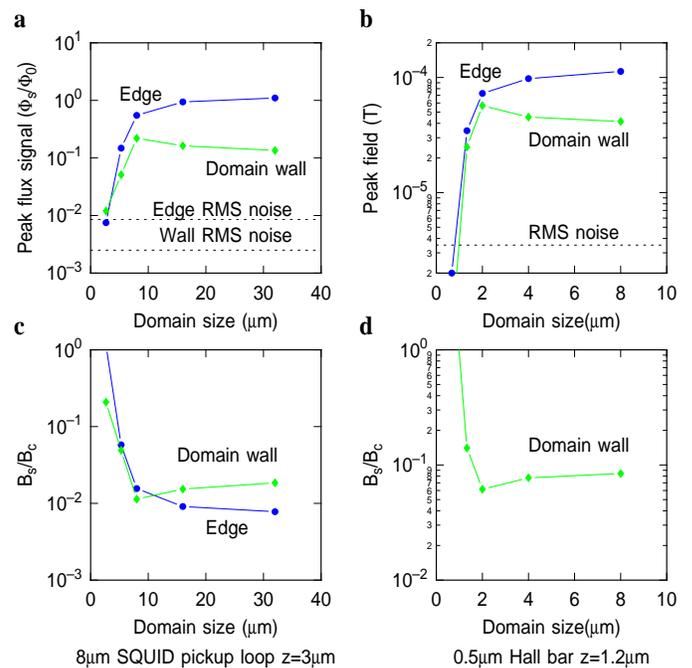}
\vspace{0.1in}
\caption{
{\bf a} Plots of the predicted peak flux signals for an 8$\mu$m square SQUID pickup loop, 3 $\mu$m above
the sample surface, for a 64 $\mu$m square $p_x \pm ip_y$ superconductor with various
domain sizes, using the predictions of Matsumoto and Sigrist \cite{matsumoto1999} for the
spontaneously generated edge and domain supercurrents. The dashed lines represent the estimated
SQUID noise in the measurements within the sample (lower line) and at the sample edges (upper line).
{\bf b} Plots of the predicted peak fields for a square Hall bar 0.5 $\mu$m on a side, 1.2$\mu$m
above the sample surface, with the corresponding Hall bar noise floor.
{\bf c} Upper limits on the size of the scaling fields
$B_s$,
normalized by $B_c= \Phi_0/2 \sqrt{2} \pi \xi_0 \lambda_L$, as a function of domain size,
given by our failure to observe
spontaneously generated supercurrents at edges and domain walls
in the SQUID measurements.
In this figure the extended-Matsumoto-Sigrist predictions are represented by $B_s/B_c$=1.
{\bf d} Upper limits on $B_s/B_c$ as a function of domain size set by the Hall bar measurements.
}
\label{fig:peakvd}
\end{figure}

Figure \ref{fig:fluxndom} shows the results from the modeling outlined above for a series of domain sizes
using parameters appropriate for our SQUID measurements.
In these calculations it was assumed that the domains were square,
and extended infinitely far in the negative $z$ direction (perpendicular to the crystal face).
Fig. \ref{fig:fluxncrs} shows cross-sections through the modeling results
as indicated by the dashed lines in Fig. \ref{fig:fluxndom}.
As expected, the magnetic fields above the edges and domain boundaries are averaged over a length set by
both the height of the sensor above the sample surface and its size.
This leads to a rapid decrease in the predicted signal
when the domains become smaller than a critical length. (In this modeling the magnetic signal for a domain size
of 4 microns vanishes everywhere except at the sample corners because, due to the symmetry of the
domains with respect to the sensor, there are exactly as many positive as negative
contributions to the flux through the 8 $\mu$m diameter pickup loop.  For this reason, we show
the predicted flux for 32/6=5.3 rather than 4 micron domains.)
The calculated peak values for the SQUID flux signal for edges and domain boundaries
are plotted in Fig. \ref{fig:peakvd}a as a function of domain size. The lower dashed line
in Fig. \ref{fig:peakvd}a
is an estimate of the noise in the SQUID images above the interior of the crystals, taken to be the rms noise
of the flux distribution shown in Fig. \ref{fig:acimage}e (2.5m$\Phi_0$). The upper dashed line
is the rms value of the flux distribution above the sample edges in Fig. \ref{fig:abimage}a (8.5m$\Phi_0$).
Comparable modeling results using parameters
appropriate for our Hall bar measurements are shown in Fig. \ref{fig:peakvd}b.
In this case the dashed line represents the rms noise value of the field distribution
in Fig. \ref{fig:hallimag}d (3.5$\mu$T).
We do not display an experimental limit on the possible
edge currents set by the Hall bar experiments
because of uncertainties associated with the hole geometry and surface damage
induced by the focussed ion beam in these experiments.\cite{bjornsson2005}

In order to place limits on the possible field magnitude and domain sizes consistent with
our results, we assume that the magnitude of the spontaneous supercurrents can vary, but that
the spatial distribution of spontaneous supercurrents is
as calculated by Matsumoto and Sigrist.
With this
assumption we can scale the results, for example, in Figs. \ref{fig:peakvd}a,b vertically
by assuming a scaling field $B_s$ different from $B_c=\Phi_0/2\sqrt{2}\pi\xi_0\lambda_L$.
In order for the spontaneous supercurrents to be unobservable in our experiments, the
scaling factor and domain size must be in the region below and to the left of
the lines in Figs. \ref{fig:peakvd}c,d.  Either the spontaneous currents
are substantially smaller than
calculated from the extended-Matsumoto-Sigrist model, or the domains are small.
For example, for the SQUID
measurements, the magnitude of the supercurrents at the edge must be
a factor of 100 smaller than those prediced by MS if the domains
are 10 or more microns in size.

\section {Surface screening effects}

In our modeling we have neglected the fact that
the magnetic fields at the sample surface will be somewhat reduced
from their bulk values.
In principle, one can calculate the surface fields
by self-consistently solving the Bogliubov-de Gennes equations in the
appropriate geometry.  Here, we simply
estimate the errors involved in
neglecting surface screening effects using a London approach.
Following Ref. \onlinecite{bluhm2007}, the superconductor is assumed to fill the half-space
$z<0$. If the change in the penetration depth close to the surface is neglected,
the magnetic field ${\bf B}$ inside the superconductor can be decomposed as ${\bf B=B_0+B_1}$,
where ${\bf B_0}$ is the particular solution given
by Matsumoto and Sigrist \cite{matsumoto1999}
of the inhomogeneous London's equation for a domain wall
and ${\bf B_1}$ is a general homogenous solution chosen
to satisfy the matching conditions at $z=0$. The London's equation for the particular solution
can be written as \cite{bluhm2007}
\begin{equation}
k(K+k)\Phi_K = K \hat{e_z} \cdot {\bf \tilde{B}}_0 ({\bf k},0)+i{\bf k} \cdot \left ({\bf \tilde{B}}_0(\bf{k},0)-4\pi
{\bf \tilde{M}}({\bf k},0) \right ),
\label{eq:phik}
\end{equation}
where $k=\sqrt{k_x^2+k_y^2}$, $K=\sqrt{k^2+1/\lambda^2}$, the magnetic field ${\bf B}$ above the superconductor
is given by ${\bf B} = - \nabla \Phi_K$,
${\bf \tilde{B}_0}$ and ${\bf \tilde{M}}$ are the
2-dimensional Fourier transforms in $x$ and $y$ of the inhomogeneous solution to London's equation and
the volume magnetization respectively, and ${\bf k} = k_x {\bf \hat{e}}_x + k_y {\bf \hat{e}}_y$.
However, if the domain walls are assumed to be parallel to the $z$ axis, both ${\bf B}_0$ and ${\bf M}$ have only
$z$ components, and Eq. (\ref{eq:phik}) reduces to
\begin{equation}
\Phi_K = \frac{K}{k(k+K)} \tilde{B}_{0z}({\bf k},0),
\label{eq:phik2}
\end{equation}
where $\tilde{B}_{0z}$ is the $z$-component of ${\bf \tilde{B}}_{0}$. Then
\begin{equation}
{\bf \tilde{B}}({\bf k},z) = \frac{(i {\bf k}+k \hat{e}_z) K}{k(k+K)} \tilde{B}_{0z}({\bf k},0)e^{-kz}.
\label{eq:b}
\end{equation}
In our case we are only interested in the $z$-component of the field outside of the superconductor, which
takes the particularly simple form
\begin{equation}
\tilde{B}_z({\bf k},z)=\frac{K}{k+K} \tilde{B}_{0z}({\bf k},0)e^{-kz}.
\label{eq:bz}
\end{equation}

\begin{figure}
\includegraphics[width=3.5in]{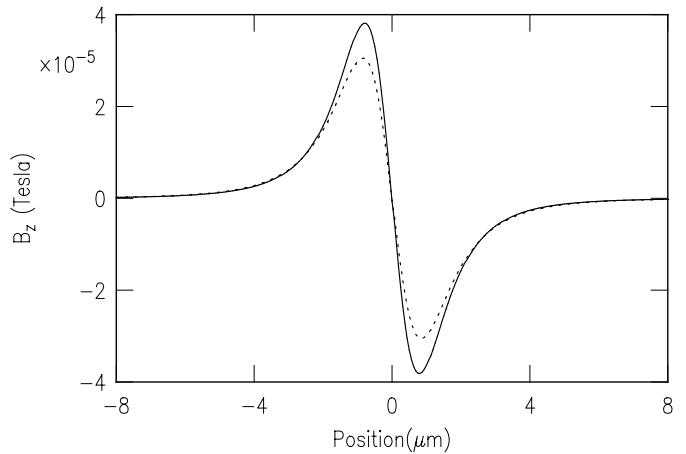}
\vspace{0.1in}
\caption{
Comparison of the predicted magnetic field sensed by a 0.5 $\mu$m square Hall bar, 1.2$\mu$m above
a single domain in a $p_x \pm ip_y$ superconductor, using the spontaneous domain currents
predicted by Matsumoto and Sigrist,\cite{matsumoto1999} with (dashed line) and without (solid line)
surface screening effects as described in the text.  The solid line corresponds to the
extended-Matsumoto-Sigrist model.
}
\label{fig:hallcrcm}
\end{figure}

The modeling in the previous section, which neglects surface shielding,
is equivalent to Eq. \ref{eq:bz} in the limit $\lambda \rightarrow 0$.
Fig. \ref{fig:hallcrcm} shows the effects of surface screening on the fields
predicted for a single domain boundary for parameters
appropriate for our Hall bar measurements.
Even in this case the effects of screening are relatively small, because the
penetration depth is smaller than the measuring height and size of the Hall bar.
Surface screening effects would be even smaller (a few percent)
for the case of SQUID imaging, because of the
larger size of the sensor. The geometry for considering the effects of superconducting
shielding on the edge fields is more complex than for the case of the domain boundary,
as one needs to consider a superconductor bounded by both $z$ and at least one of
$x$ or $y$.  However, again the edge and surface effects will be confined
on the scale of the penetration depth which is much smaller than the distance to
the probe or the probe size.  Therefore,
we do not believe that the simple model presented above will be more than a factor
of two different from a full calculation.

\section{Discussion}

If the superconductivity of Sr$_2$RuO$_4$ breaks time-reversal symmetry,
it should spontaneously generate supercurrents at domain boundaries and
sample edges.
The fact that no  magnetic fields due to such
supercurrents were observed
using scanning magnetic microscopy places significant limits
on the size of these currents and the size of the domains, as shown
in Fig. \ref{fig:peakvd}.  In particular, from
the combined Hall bar and SQUID measurements, we conclude that if the
spontaneous supercurrents at a domain wall are of the size expected from
the calculations of Matsumoto and Sigrist\cite{matsumoto1999} and the modelling done
here, one can set a conservative upper limit on the domain size of 1.5 microns
for both interior and edge domains.
Alternatively,
if the domains intersecting the ab face are 10 microns or more in size, we
conclude that the spontaneous supercurrents at edges are
a factor of 100 smaller than expected from the calculations of Matsumoto and
Sigrist combined with our modelling.

Calculations of the self-consistent screening currents employed
in our modelling have assumed an ideal $p_x\pm ip_y$ superconducting
gap symmetry.\cite{matsumoto1999}  However,
for Sr$_2$RuO$_4$, the gap in the $ab$ plane is believed to
be anisotropic.\cite{deguchi2004a,deguchi2004b}  In addition,
three different bands contribute to the Fermi surface in
Sr$_2$RuO$_4$.\cite{mackenzie2003}
These properties are likely to impact on the magnitude of the self-consistent
screening currents, although, a priori, it is not clear whether the magnitude
would be increased or decreased from the values calculated by MS.
On the other hand, muon spin resonance observed internal fields which
are roughly consistent with the predicted values.\cite{luke1998}  If these
observed fields are due to internal domain walls, it suggests that the
surface currents must be reduced by two or more orders of magnitude
from their bulk values if the domains are larger than 10 microns.
It is difficult to imagine what could so strongly reduce the surface fields
at the ab surface due to domain boundaries.  The surface screening
effects are small; the surfaces are cleaved; and roughness even to the
depth of a hundred angstroms or so will not substantially reduce the fields
detected at the Hall probe or SQUID.
This suggests that domains intersecting the ab plane are either so sparse
as to not have been scanned or are smaller than a few microns.
Another possibility is that the domains intersecting the ab surface are
shallow, with a depth along the c-axis noticeably less than the
penetration depth.  In this case, the spontaneous currents and fields
could be too weak and spread out in the ab layers to be detected.
However, we note that either small domains or domains shallower than
the optical skin depth
would also interfere with observations of the Kerr effect rotation.

Sufficient roughness of the ac or bc faces can be expected to have a more noticeable
effect on the edge or boundary currents.  The samples used for the SQUID
measurements on these faces were polished and AFM imaging on typical
samples show them to be smooth to 5 nm (rms).\cite{karlthesis}
MS assumed specular scattering from the edge in which case one component
of the order parameter is suppressed while the other component is
slightly enhanced.  For diffuse scattering from a rough edge, both
components will be suppressed and this will reduce the surface
currents and resulting magnetic fields.  Although self-consistent
calculations have not been carried out for this case, the effect
of surface roughness on the two component order parameter has been
studied,\cite{nagato98} and one finds that the two components heal
over quite different length scales.  Using Ginzburg-Landau and
London theory to estimate
the resulting change in the surface magnetization, one finds that,
even for completely diffuse scattering from a rough surface, the
reduction in surface magnetization is less than 30\%.\cite{kallin}

Domain walls cost energy because they disrupt the superconducting order.
Unlike a ferromagnet, there is no balancing of this energy due to
dipolar forces because spontaneous screening currents ensure that the
magnetic field, or local magnetization, is zero inside the superconductor.
Therefore, in principle, a single domain $p_x + ip_y$ superconductor
is possible.  However, domains will naturally form as the sample is
cooled through $T_c$ and as extended objects, these domains are susceptible
to pinning by defects and impurities in the sample.  Therefore, one
expects domains to be present although their density may be controlled
by sample purity and slow cooling in a field.  Muon spin resonance
experiments were interpreted as evidence for dilute domains,\cite{luke1998}
Kerr effect measurements suggested domain sizes in the range of 50 to 100
microns,\cite{xia2006} while the first phase sensitive Josephson tunneling measurements
are consistent with no domains
(or a small even number of domain walls between opposing faces of the crystal).\cite{nelson2004}
On the other hand, more recent
Josephson tunneling measurements were interpreted as evidence for
dynamic domains of $\sim$1 micron on average,\cite{kidwingara2006}
although one would extract
larger domain sizes if finite domains perpendicular to the c-axis were
included in the modelling.  All of these measurements, except for muon
spin resonance, would see reduced signals if the domain size along the
c-axis becomes small and this would affect the measurements reported here
as well.  Unless the fields at domain walls are reduced by more than an
order of magnitude in size from the predicted values, the Hall bar
measurements suggest domain sizes of either less than 1.5 microns in size
over the ab face or large enough that no domain wall fell in the
100$\times$100 $\mu$m scan area.

Earlier work has reported that large domains can be flipped by fields
of the order of a mT or larger\cite{xia2006} and that small surface domains are influenced
by fields $< 0.1\mu$T.\cite{kidwingara2006}
While the data presented here was
taken on samples cooled in fields less than 2.5$\mu$T, Hall
data taken on samples cooled in up to a mT was very similar
to that shown here except for the presence of more trapped vortices.\cite{bjornsson2005}
In principle, very fast domain wall motion could result in zero time-averaged edge current
and zero time-averaged domain wall current.  However, previous
experiments\cite{nelson2004,kidwingara2006,xia2006}
suggest that the domain wall motion would be slow on our experimental time scale, which is 10 secconds
per line scan for the scanning SQUID microscope data shown here. Therefore, it is
unlikely that dynamic behavior of the domains prevents the observation of the signal in this experiment.

In conclusion, scanning magnetic microscopy measurements place quite
severe limits on the size of edge currents and/or on domain sizes in
Sr$_2$RuO$_4$.  The different experimental results taken as evidence for
 $p_x + ip_y$ pairing come to quite different conclusions about domain
sizes.  Since there are now detailed predictions for the field profile
in the vicinity of domain walls in the bulk, muon spin resonance could
now, in principle, provide detailed information about the validity of
these predictions as well as quantitative information
about the density of domains in the bulk.
In addition,
either slow muons\cite{sigrist} or beta-NMR\cite{salman2006}
could be used to probe the
surface region and to look
for fields due to spontaneous edge currents as well as domains near the
surface. Scanning
magnetic microscopy is still one of the most direct probes of domains
intersecting the surface and of edge currents
and further improvements in sensitivity may
either confirm or rule out their existence.

\section{Acknowledgements}

We would like to thank
H. Bluhm,
J. Berlinsky,
S.B. Chung,
M. Franz,
H. Hilgenkamp,
T.-L. Ho,
M. Matsumoto,
M. Sigrist
and
M. Stone
for many useful
discussions and H. Bluhm for also sharing the results
of his calculations before publication. We would also like to thank P. Bj{\"o}rnsson
for sharing his Hall bar data with us.
This work was supported by DOE at Stanford under grant
DE-AC02-76SF00515.
Work at Penn State was supported by DOE under grant DE-FG02-04ER46159.
JRK was supported by
the Center for Probing the Nanoscale,
an NSF NSEC, NSF Grant No. PHY-0425897, and by
the Dutch NWO Foundation.
In addition, CK acknowledges the hospitality of
the Stanford Institute for Theoretical Physics and the Kavli
Institute for Theoretical Physics during this collaboration and the support
of the National Science Foundation under Grant No. PHY05-51164.

\end{document}